\documentclass{webofc}

\usepackage[varg]{txfonts}   
\usepackage{hyperref}
\usepackage{url}
\hypersetup{colorlinks=true,citecolor=blue,urlcolor=blue,linkcolor=blue}
\begin{document}
\title{Enhancing software-hardware co-design for HEP by low-overhead profiling of single- and multi-threaded programs on diverse architectures with Adaptyst}

\author{\firstname{Maksymilian} \lastname{Graczyk}\inst{1}\fnsep\thanks{Corresponding author: maksymilian.graczyk@cern.ch} \and
        \firstname{Stefan} \lastname{Roiser}\inst{1}
}

\institute{European Organisation for Nuclear Research (CERN), Geneva, Switzerland}

\abstract{
Given the recent technological trends and novel computing paradigms spanning both software and hardware, physicists and software developers can no longer just rely on computers becoming faster to meet the ever-increasing computing demands of their research. Adapting systems to the new environment may be difficult though, especially in case of large and complex applications. Therefore, we introduce Adaptyst (formerly AdaptivePerf): an open-source and architecture-agnostic tool aiming for making these computational and procurement challenges easier to address. At the moment, Adaptyst profiles on- and off-CPU activity of codes, traces all threads and processes spawned by them, and analyses low-level software-hardware interactions to the extent supported by hardware. The tool addresses the main shortcomings of Linux "perf" and has been successfully tested on x86-64, arm64, and RISC-V instruction set architectures. Adaptyst is planned to be evolved towards a software-hardware co-design framework which scales from embedded to high-performance computing in both legacy and new applications and takes into account a bigger picture than merely choosing between CPUs and GPUs. Our paper describes the current development of the project and its roadmap.
}

\maketitle

\section{Introduction}
\label{intro}
The sheer complexity of high energy physics (HEP) experiments and accelerators creates several computing challenges. For example, the Large Hadron Collider (LHC) at CERN is the most powerful particle accelerator in the world, where an order of one billion particle collisions is produced every second \cite{lhc}. Managing the machinery, preparing experiments, and processing the resulting data are computationally-expensive tasks that need to be done at all levels, from various accelerator controls, through trigger and data acquisition (DAQ) systems, up to data centre, grid, and high-performance computing (HPC) setups.
\\\\
At the same time, the computing technology landscape is rapidly changing: Moore's law and Dennard scaling are slowing down, the awareness of the need for sustainable computing is increasing, more and more specialised hardware is arriving, the focus on utilising such devices is growing, and novel paradigms spanning both software and hardware are appearing. The recent examples are GPUs used for accelerating large language model training, FPGAs being deployed to more and more HPC centres along with being used in real-time systems, as well as custom accelerators such as ones based on RISC-V.
\\\\
Under these circumstances and with the number of collisions at the LHC alone due to increase by a factor of 5 to 7.5 by 2030s through the High Luminosity LHC (HL-LHC) upgrade \cite{hllhc}, relying on computers becoming faster or optimising the existing software only algorithmically may no longer be sufficient to meet the growing requirements of HEP research. However, adapting HEP applications to the new technological trends is not straightforward, especially in case of large and complex implementations. This is where automated performance analysis and profiling can bridge the gap, but finding a suitable tool is challenging when wide architectural support and reliability are important.
\\\\
To tackle this problem, we introduce Adaptyst (formerly AdaptivePerf): an open-source and architecture-agnostic tool aiming for helping physicists, software/hardware engineers, and administrators address their computational performance and procurement challenges more easily. At its current development stage, Adaptyst profiles on- and off-CPU activity of codes running on Linux using our patched version of "perf" \cite{customperf}, traces all the threads and processes spawned by them, and analyses low-level software-hardware interactions to the extent supported by hardware. The tool has been successfully tested on machines based on x86-64, arm64, and RISC-V instruction set architectures (ISAs).
\\\\
This paper presents the current early development of Adaptyst and its roadmap. At the moment, our work addresses some of the main shortcomings of upstream "perf" \cite{perf} for programs compiled with frame pointers, such as limited off-CPU profiling and incomplete stack traces without an apparent cause. Eventually, Adaptyst will deploy several methods in addition to profiling and be a software-hardware co-design framework scaling from embedded/bare-metal/edge computing to HPC. Compared to the prior work, our tool will:
\begin{enumerate}
\item extend easily to all present and future workflow and system/hardware types;
\item work with legacy codes and across programming languages with no explicit accelerated code separation required;
\item perform automated software-hardware co-design with a complete macroscopic picture view rather than with a narrow focus on deciding only where hardware acceleration should be done.
\end{enumerate}

\noindent
In section \ref{adaptyst}, we describe in detail how the tool works and how its results can be analysed effortlessly with our web analysis program called Adaptyst Analyser (formerly AdaptivePerfHTML). In section \ref{future}, we discuss the upcoming evolution of the project towards a comprehensive software-hardware co-design tool for HEP and beyond. Afterwards, we examine the existing work in section \ref{related-work} and conclude our paper in section \ref{conclusions}.

\section{Performance analysis with Adaptyst and Adaptyst Analyser}
\label{adaptyst}
\begin{figure}[h]
\centering
\includegraphics[width=\textwidth,clip]{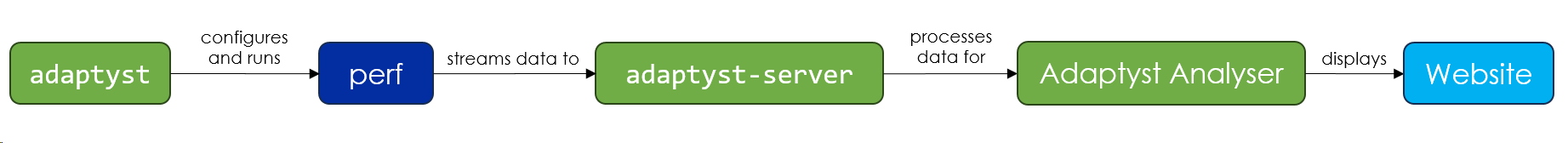}
\caption{The full program flow of Adaptyst, starting with the \texttt{adaptyst} command run by a user. \texttt{adaptyst-server} is a component of Adaptyst launched by Adaptyst in the background by default.}
\label{adaptyst-working}
\end{figure}

\begin{figure}[h]
\centering
\includegraphics[width=\textwidth,clip]{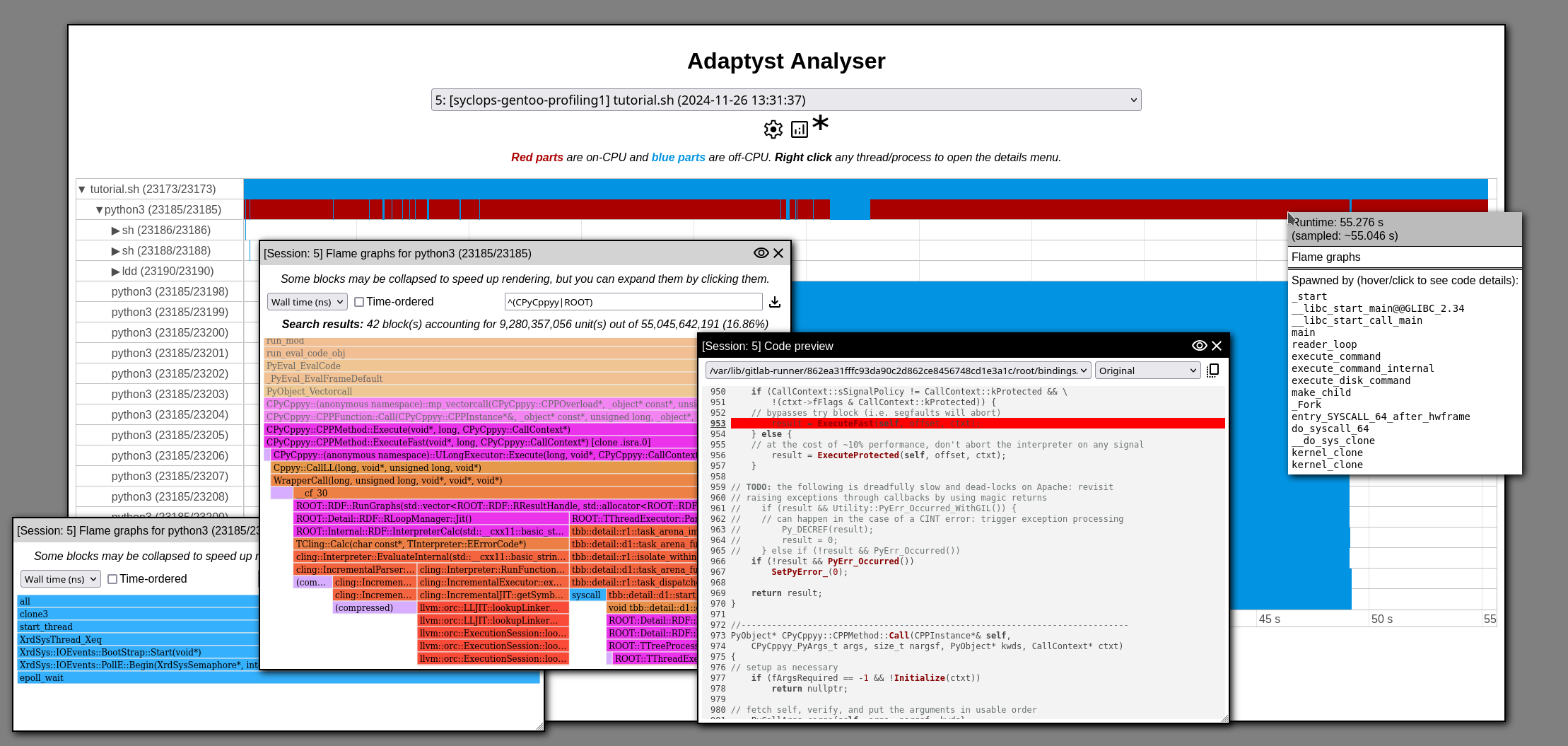}
\caption{A profiling session for a CERN ROOT \cite{brun1997root} example opened in Adaptyst Analyser. The "General analyses" icon is marked with an asterisk (*).}
\label{adaptyst-analyser-main}
\end{figure}

At present, Adaptyst is an open-source, language-agnostic, and architecture-agnostic code profiler for Linux. It is written mostly in C++ and Python and built on top of "perf" \nolinebreak \cite{perf} with the custom patches \cite{customperf}. For any given command, the tool samples both on-CPU and off-CPU time in code detail, traces every spawned thread and process, and minimises the risk of broken profiled stacks by detecting inappropriate kernel and CPU configurations automatically (such as the wrong value of the \texttt{kernel.perf\_event\_max\_stack} setting or the activated NUMA memory balancing mechanism).
\\
The main functionality of Adaptyst is designed with hardware and vendor portability in mind and has been successfully tested on x86-64, arm64, and RISC-V machines. If profiling low-level software-hardware interactions is desired (e.g. page faults, cache misses, retired instructions), our work supports it to the extent allowed by hardware.
\\\\
Installing and getting started with Adaptyst is straightforward: the early development version is available on GitHub\footnote{\url{https://github.com/adaptyst/adaptyst}. The latest commit referred to by this paper is \texttt{8abc18c}.} with various setup options. Commands can be profiled by executing \texttt{adaptyst <optional options> ---- <command>}. As shown in figure \ref{adaptyst-working}, when the \texttt{adaptyst} command is run, it configures and runs "perf" under the hood, which then profiles and streams events on-the-fly to \texttt{adaptyst-server} started in the background. \texttt{adaptyst-server} is an Adaptyst component responsible for processing profiling data into a format that can be later analysed  with e.g. Adaptyst Analyser.
\\
While \texttt{adaptyst-server} is launched in the background automatically by Adaptyst by default, it can also be set up manually on a separate machine from the one running the profiled program. The data transmission is then done by TCP, allowing moving the resource-expensive part of profiling to a (potentially) more computationally-powerful computer and avoiding any interference between profiling and the profiled program. Because all the information necessary for later analysis is gathered by Adaptyst, \texttt{adaptyst-server} does not require any profiled executables or debug symbols and can run on a non-Linux machine.
\\\\
After Adaptyst completes its job, the results are produced in a non-encrypted format that can be analysed in a variety of ways, e.g. with Adaptyst Analyser. Adaptyst Analyser is an open-source program (the early development version is available on GitHub\footnote{\url{https://github.com/adaptyst/adaptyst-analyser}. The latest commit referred to by this paper is \texttt{e6277ee}.}) which renders an interactive website displaying all the information gathered in an Adaptyst profiling session. It can be run using the command \texttt{adaptyst-analyser <optional options> <path to results>}, which returns the address to open in a web browser.
\\\\
On the website, a selected session is summarised by the timeline view containing the thread/process hierarchy tree on the left and on-CPU/off-CPU activity on the right, where the red parts correspond to on-CPU time and the blue parts correspond to off-CPU time. Every thread/process can be inspected in more detail by right-clicking it and choosing a desired analysis type to open, with the spawning stack trace shown at the same time if available. Multiple analyses can be displayed simultaneously from the same or different sessions, allowing direct side-by-side comparisons. Figure \ref{adaptyst-analyser-main} presents a sample session opened in Adaptyst Analyser.
\\\\
In the current version of Adaptyst, a detailed activity of any given thread/process is visualised with flame graphs \cite{flamegraph}: one hot-and-cold flame graph for overall on-CPU and off-CPU runtime and one flame graph per low-level software-hardware interaction type if selected by the user during profiling. These are interactive: they can be zoomed in/out, ordered chronologically, and queried with search phrases (regular expressions are supported). If a source code is available, it can be opened within Adaptyst Analyser with the line breakdown of most resource-consuming parts by right-clicking a flame graph block of interest, as shown in the "Code preview" window in figure \ref{adaptyst-analyser-main}. The code can also be opened for spawning stack traces, with the specific spawning line highlighted.
\\\\
Additionally, Adaptyst Analyser supports displaying cache-aware roofline plots produced by the CARM tool \cite{morgado2024carm}. If the roofline benchmarks have been executed and saved alongside Adaptyst results, the plot can be opened by clicking the "General analyses" icon marked with an asterisk in figure \ref{adaptyst-analyser-main} and selecting "Cache-aware roofline model". Plotting flame graph blocks of interest as single points on the plot will be added in the future.

\section{Roadmap towards software-hardware co-design}
\label{future}
\begin{figure}[h]
\centering
\includegraphics[width=10cm,clip]{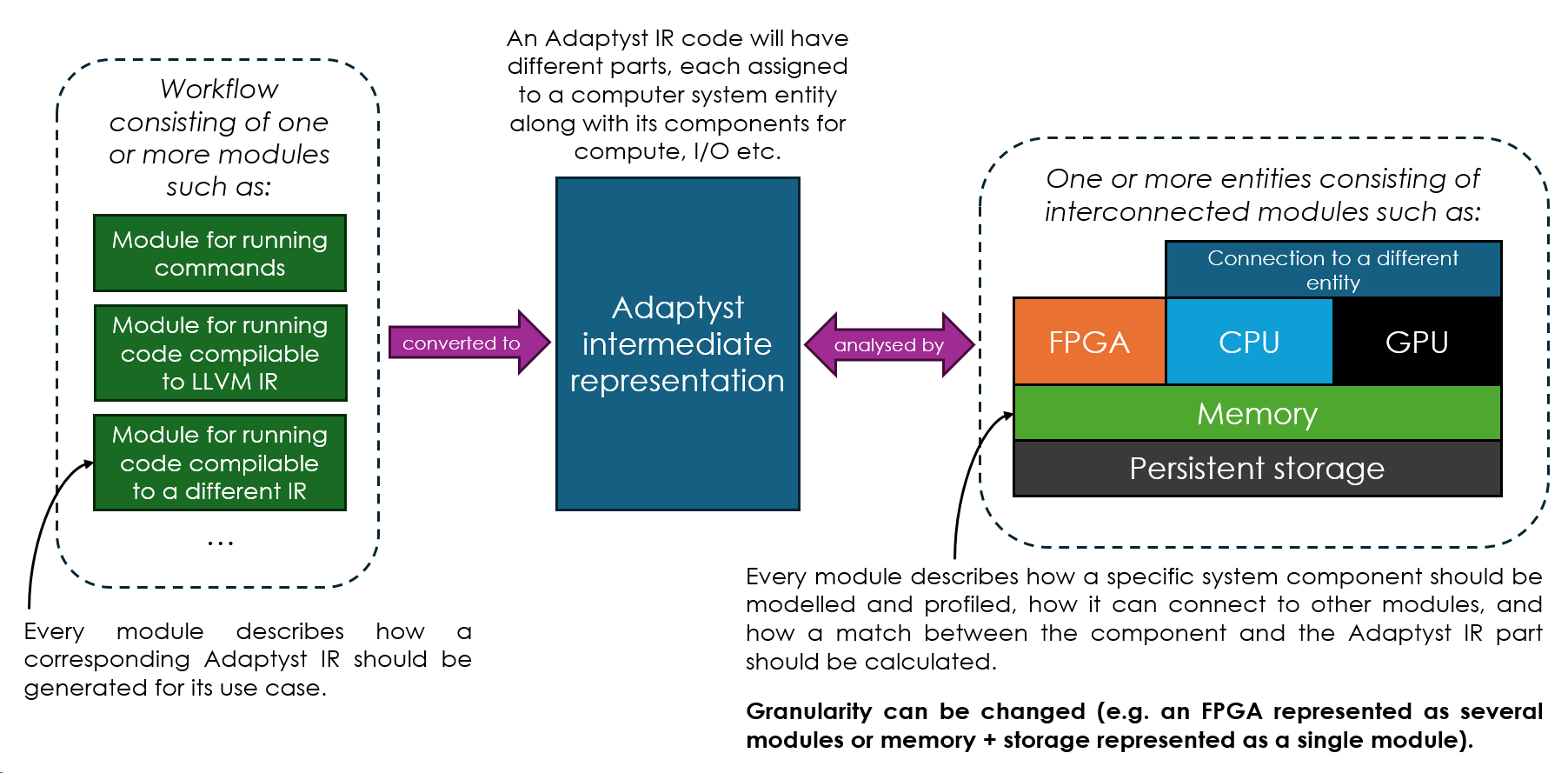}
\caption{The proposed modular design of Adaptyst as a software-hardware co-design framework.}
\label{adaptyst-future}
\end{figure}

While Adaptyst currently serves as a profiling tool, the goal of the project is delivering a single and unified platform for comparing and customising off-the-shelf and in-house software and hardware architectures across the entire spectrum, from embedded/bare-metal/edge computing to HPC. By doing so, virtually every computing challenge faced by HEP research and other fields will be covered.
\\\\
To achieve this goal, we plan to decouple Adaptyst from "perf" (including by replacing the current patched-"perf" arrangement with an in-house code using e.g. the \texttt{perf\_event\_open(2)} system call and eBPF), develop an Adaptyst intermediate representation (IR), and build the entire Adaptyst suite around it in a modular fashion as explained in figure \ref{adaptyst-future}. The user would provide their workflow to be analysed as an input in form of user-configured workflow modules. These would be created by external developers such as compiler engineers and would describe how a corresponding IR should be generated for their use case (e.g. a module for running commands or a module for running code compilable to LLVM IR \cite{lattner2004llvm}). As the result, Adaptyst would carry out comprehensive performance analysis and suggest the best system and hardware combination given the user requirements and constraints, including elements such as CPUs, GPUs, FPGAs, memory hierarchy, storage solutions, networking, etc.
\\
Hardware-related analysis would be performed with the help of system component modules developed by external developers such as hardware and performance engineers. They would describe how a specific system component should be modelled and profiled, how it can connect to other system modules, and how a match between the component and the IR part should be determined quantitatively for a given category (e.g. latency, power consumption). Their granularity can be changed, so e.g. compared to figure \ref{adaptyst-future}, an FPGA can be represented as several modules and memory and storage can be represented together as a single module.
\\\\
The most optimal software-hardware co-design given the user constraints would be obtained by maximising all matches of interest. This would also include automatic system component selection and automatic IR part splitting so that various system and hardware choices can be evaluated and no explicit accelerated code separation is required.
\\\\
The current Adaptyst work described in this paper would fit within this roadmap by having a workflow module for running arbitrary commands and a system component module for analysing CPU activity of a given code on Linux. The modular design would allow adding support for other workflows and systems without having to modify Adaptyst itself.
\section{Related work}
\label{related-work}
The customisation target in software-hardware co-design is fluid: it can vary between software and hardware. On the software side, apart from standard compilers such as GCC, there are domain-specific languages and transparent compilers (e.g. Liquid Metal \cite{auerbach2012compiler}, Dandelion \cite{rossbach2013dandelion}, MATCH \cite{banerjee2000matlab}, Polly-ACC \cite{grosser2016polly}, SPIRAL \cite{franchetti2018spiral}, HTrOP \cite{riebler2019transparent}), data layout abstractions (e.g. LLAMA \cite{gruber2023llama}) and various parallel programming models which can be vendor-specific (e.g. NVIDIA CUDA) or -agnostic (e.g. SYCL \cite{sycl}, Kokkos \cite{trott2021kokkos}, Alpaka \cite{zenker2016alpaka}, OpenACC \cite{openacc}). All of these try to facilitate performance programming and programming heterogeneous/parallel platforms efficiently for software developers. While they have significant promise in terms of code portability except for the vendor-specific models, their adoption is hindered by one or more of these issues: no hardware optimisations, an explicit accelerated code needed (potentially with a different paradigm than the rest of the code), uncertain long-term support, limited language support, limited automatic hardware or accelerated code selection, and a narrow focus on parallelisation only.
\\\\
On the hardware side, there is high-level synthesis (HLS) generating a hardware implementation of a software code (e.g. Vitis HLS \cite{vitishls}, Catapult HLS \cite{catapulthls}, Dynamatic \cite{josipovic2021synthesizing}). While this solves the problem of no hardware optimisations, there is an inherent assumption that custom hardware must be generated for an entire provided code. This creates suboptimal implementations for programs most suitable for standard CPUs for example. Also, for best performance, high-level synthesis codes should be written in a more VHDL-/Verilog-like way despite using typical high-level software languages such as C++. Domain-specific wrappers address this issue though (e.g. hls4ml \cite{duarte2018fast}). Moreover, HLS has the similar lack of the bigger picture and limited language support, making this not a universal solution.
\\\\
In-between, there are modular compilation frameworks like LLVM \cite{lattner2004llvm}, MLIR \cite{lattner2021mlir}, DaCe \cite{ben2019stateful}. They strive to have the best of both worlds: effortless programming from the developer point of view, the potential for optimising both software and hardware, and better scalability across languages. However, these frameworks also suffer from the lack of the bigger picture and limited automatic hardware selection / accelerated code selection. Adaptyst will bridge the gap here by solving these problems and allowing varying the customisation target based on user requirements and constraints.
\\\\
In terms of profiling, there are several established and actively-maintained tools that are feature-rich and state-of-the-art for the hardware they target, e.g. upstream "perf" \cite{perf}, Intel VTune Profiler \cite{vtune}, AMD \textmu Prof \cite{uprof}, valgrind \cite{valgrind}, gprof \cite{gprof}, gperftools \cite{gperftools}, NVIDIA Nsight \cite{nsight}, TAU \cite{shende2006tau}, and HPCToolkit \cite{adhianto2010hpctoolkit}. However, none of these programs offer flexible support of heterogeneous and custom architectures to the extent planned to be implemented in Adaptyst. Intel VTune Profiler, AMD \textmu Prof, and NVIDIA Nsight are also proprietary and not vendor-portable unlike our tool. TAU and HPCToolkit do not run on RISC-V either.
\\\\
It is important to note that Adaptyst aims for interacting with the existing work in both software-hardware co-design and profiling, rather than competing with it.

\section{Conclusions}
\label{conclusions}
This paper presents Adaptyst: an open-source, language-agnostic, and architecture-agnostic tool helping physicists, software/hardware engineers, and administrators address their computing and procurement needs more easily. The project is in its early development stage, where the current main functionality is profiling codes running on Linux. Adaptyst will later be evolved towards a comprehensive software-hardware co-design framework scaling across all use cases in HEP and beyond, working for both legacy and new applications, and taking the big picture into consideration such as storage and networking solutions.
\\\\
The current main limitations of our work are Linux-only support and requiring codes to be compiled with frame pointers for obtaining complete stack traces. These shortcomings will be addressed in the future as part of the roadmap described in section \ref{future}, e.g. by adding support for applications running on bare-metal hardware with the help of an appropriate system component module.
\\\\
Given the rapidly-changing technological landscape, the rising complexity of computer systems, and growing computational requirements of fundamental research and other fields, Adaptyst has the potential for decreasing the effort needed to adapt software to this new environment and therefore helping achieve breakthroughs more easily, rapidly, and sustainably.

\section*{Acknowledgements}
This work is supported by the European Union HE research and innovation programme, grant agreement No 101092877 (SYCLOPS).

\bibliography{bibliography.bib}

\begin{thebibliography}{33}

\bibitem{lhc}
CERN, Facts and figures about the lhc, \url{https://home.web.cern.ch/resources/faqs/facts-and-figures-about-lhc}, access: 2025-02-23

\bibitem{hllhc}
CERN, High-luminosity lhc, \url{https://home.cern/resources/faqs/high-luminosity-lhc}, access: 2025-02-23

\bibitem{customperf}
Patched "perf" repository on cern gitlab, \url{https://gitlab.cern.ch/adaptyst/linux}, access: 2025-02-27

\bibitem{perf}
perf: Linux profiling with performance counters, \url{https://perfwiki.github.io/main}, access: 2025-02-23

\bibitem{brun1997root}
R.~Brun, F.~Rademakers, Root—an object oriented data analysis framework, Nuclear instruments and methods in physics research section A: accelerators, spectrometers, detectors and associated equipment \textbf{389}, 81 (1997).

\bibitem{flamegraph}
B.~Gregg, Flame graphs, \url{https://brendangregg.com/flamegraphs.html}, access: 2025-02-23

\bibitem{morgado2024carm}
J.~Morgado, L.~Sousa, A.~Ilic, CARM Tool: Cache-Aware Roofline Model Automatic Benchmarking and Application Analysis, in \emph{2024 IEEE International Symposium on Workload Characterization (IISWC)} (IEEE, 2024), pp. 68--81

\bibitem{lattner2004llvm}
C.~Lattner, V.~Adve, LLVM: A compilation framework for lifelong program analysis \& transformation, in \emph{International symposium on code generation and optimization, 2004. CGO 2004.} (IEEE, 2004), pp. 75--86

\bibitem{auerbach2012compiler}
J.~Auerbach, D.F. Bacon, I.~Burcea, P.~Cheng, S.J. Fink, R.~Rabbah, S.~Shukla, A compiler and runtime for heterogeneous computing, in \emph{Proceedings of the 49th Annual Design Automation Conference} (2012), pp. 271--276

\bibitem{rossbach2013dandelion}
C.J. Rossbach, Y.~Yu, J.~Currey, J.P. Martin, D.~Fetterly, Dandelion: a compiler and runtime for heterogeneous systems, in \emph{Proceedings of the Twenty-Fourth ACM Symposium on Operating Systems Principles} (2013), pp. 49--68

\bibitem{banerjee2000matlab}
P.~Banerjee, N.~Shenoy, A.~Choudhary, S.~Hauck, C.~Bachmann, M.~Haldar, P.~Joisha, A.~Jones, A.~Kanhare, A.~Nayak et~al., A MATLAB compiler for distributed, heterogeneous, reconfigurable computing systems, in \emph{Proceedings 2000 IEEE Symposium on Field-Programmable Custom Computing Machines (Cat. No. PR00871)} (IEEE, 2000), pp. 39--48

\bibitem{grosser2016polly}
T.~Grosser, T.~Hoefler, Polly-ACC transparent compilation to heterogeneous hardware, in \emph{Proceedings of the 2016 International Conference on Supercomputing} (2016), pp. 1--13

\bibitem{franchetti2018spiral}
F.~Franchetti, T.M. Low, D.T. Popovici, R.M. Veras, D.G. Spampinato, J.R. Johnson, M.~P{\"u}schel, J.C. Hoe, J.M. Moura, Spiral: Extreme performance portability, Proceedings of the IEEE \textbf{106}, 1935 (2018).

\bibitem{riebler2019transparent}
H.~Riebler, G.~Vaz, T.~Kenter, C.~Plessl, Transparent acceleration for heterogeneous platforms with compilation to opencl, ACM Transactions on Architecture and Code Optimization (TACO) \textbf{16}, 1 (2019).

\bibitem{gruber2023llama}
B.M. Gruber, G.~Amadio, J.~Blomer, A.~Matthes, R.~Widera, M.~Bussmann, Llama: The low-level abstraction for memory access, Software: Practice and Experience \textbf{53}, 115 (2023).

\bibitem{sycl}
{Khronos Group}, Sycl overview, \url{https://www.khronos.org/sycl}, access: 2025-02-23

\bibitem{trott2021kokkos}
C.R. Trott, D.~Lebrun-Grandi{\'e}, D.~Arndt, J.~Ciesko, V.~Dang, N.~Ellingwood, R.~Gayatri, E.~Harvey, D.S. Hollman, D.~Ibanez et~al., Kokkos 3: Programming model extensions for the exascale era, IEEE Transactions on Parallel and Distributed Systems \textbf{33}, 805 (2021).

\bibitem{zenker2016alpaka}
E.~Zenker, B.~Worpitz, R.~Widera, A.~Huebl, G.~Juckeland, A.~Kn{\"u}pfer, W.E. Nagel, M.~Bussmann, Alpaka--an abstraction library for parallel kernel acceleration, in \emph{2016 IEEE International Parallel and Distributed Processing Symposium Workshops (IPDPSW)} (IEEE, 2016), pp. 631--640

\bibitem{openacc}
Openacc, \url{https://www.openacc.org}, access: 2025-02-23

\bibitem{vitishls}
AMD, Vitis hls, \url{https://www.amd.com/en/products/software/adaptive-socs-and-fpgas/vitis/vitis-hls.html}, access: 2025-02-23

\bibitem{catapulthls}
Siemens, Catapult high-level synthesis \& verification, \url{https://eda.sw.siemens.com/en-US/ic/catapult-high-level-synthesis}, access: 2025-02-23

\bibitem{josipovic2021synthesizing}
L.~Josipovic, A.~Guerrieri, P.~Ienne, Synthesizing general-purpose code into dynamically scheduled circuits, IEEE Circuits and Systems Magazine \textbf{21}, 97 (2021).

\bibitem{duarte2018fast}
J.~Duarte, S.~Han, P.~Harris, S.~Jindariani, E.~Kreinar, B.~Kreis, J.~Ngadiuba, M.~Pierini, R.~Rivera, N.~Tran et~al., Fast inference of deep neural networks in fpgas for particle physics, Journal of instrumentation \textbf{13}, P07027 (2018).

\bibitem{lattner2021mlir}
C.~Lattner, M.~Amini, U.~Bondhugula, A.~Cohen, A.~Davis, J.~Pienaar, R.~Riddle, T.~Shpeisman, N.~Vasilache, O.~Zinenko, MLIR: Scaling compiler infrastructure for domain specific computation, in \emph{2021 IEEE/ACM International Symposium on Code Generation and Optimization (CGO)} (IEEE, 2021), pp. 2--14

\bibitem{ben2019stateful}
T.~Ben-Nun, J.~de~Fine~Licht, A.N. Ziogas, T.~Schneider, T.~Hoefler, Stateful dataflow multigraphs: A data-centric model for performance portability on heterogeneous architectures, in \emph{Proceedings of the International Conference for High Performance Computing, Networking, Storage and Analysis} (2019), pp. 1--14

\bibitem{vtune}
Intel, Fix performance bottlenecks with intel vtune profiler, \url{https://www.intel.com/content/www/us/en/developer/tools/oneapi/vtune-profiler.html}, access: 2025-02-23

\bibitem{uprof}
AMD, Amd \textmu prof, \url{https://www.amd.com/en/developer/uprof.html}, access: 2025-02-23

\bibitem{valgrind}
Valgrind home, \url{https://valgrind.org}, access: 2025-02-23

\bibitem{gprof}
Gnu gprof, \url{https://ftp.gnu.org/old-gnu/Manuals/gprof-2.9.1/html_mono/gprof.html}, access: 2025-02-23

\bibitem{gperftools}
gperftools repository, \url{https://github.com/gperftools/gperftools}, access: 2025-02-23

\bibitem{nsight}
NVIDIA, Nvidia nsight developer tools - nvidia docs, \url{https://docs.nvidia.com/nsight-developer-tools/index.html}, access: 2025-02-23

\bibitem{shende2006tau}
S.S. Shende, A.D. Malony, The tau parallel performance system, The International Journal of High Performance Computing Applications \textbf{20}, 287 (2006).

\bibitem{adhianto2010hpctoolkit}
L.~Adhianto, S.~Banerjee, M.~Fagan, M.~Krentel, G.~Marin, J.~Mellor-Crummey, N.R. Tallent, Hpctoolkit: Tools for performance analysis of optimized parallel programs, Concurrency and Computation: Practice and Experience \textbf{22}, 685 (2010).

\end{thebibliography}

\end{document}